\documentclass[conference]{IEEEtran}
\IEEEoverridecommandlockouts
\usepackage{cite}
\usepackage{amsmath,amssymb,amsfonts}
\usepackage{algorithmic}
\usepackage{graphicx}
\usepackage{textcomp}
\usepackage{xcolor}

\usepackage{float}
\usepackage{booktabs}
\usepackage{makecell}
\usepackage{multirow}
\usepackage{bbding}
\usepackage{pifont}
\usepackage{wasysym}
\usepackage{utfsym}
\usepackage{fontawesome}
\def\BibTeX{{\rm B\kern-.05em{\sc i\kern-.025em b}\kern-.08em
    T\kern-.1667em\lower.7ex\hbox{E}\kern-.125emX}}
\begin{document}

\title{MFHCA: Enhancing Speech Emotion Recognition Via Multi-Spatial Fusion and Hierarchical Cooperative Attention}

\author{\IEEEauthorblockN{Xinxin Jiao}
\IEEEauthorblockA{\textit{School of Computer Science} \\
\textit{and Technology}\\
\textit{Xinjiang University}\\
Urumqi, China \\
jxx2022@xju.edu.cn}
\and
\IEEEauthorblockN{Liejun Wang$^{1\dagger}$\thanks{$^{\dagger}$ Corresponding author.}}
\IEEEauthorblockA{\textit{School of Computer Science} \\
\textit{and Technology}\\
\textit{Xinjiang University}\\
Urumqi, China \\
wljxju@xju.edu.cn}
\and
\IEEEauthorblockN{Yinfeng Yu}
\IEEEauthorblockA{\textit{School of Computer Science} \\
\textit{and Technology}\\
\textit{Xinjiang University}\\
Urumqi, China \\
yuyinfeng@xju.edu.cn}

}

\maketitle

\begin{abstract}
Speech emotion recognition is crucial in human-computer interaction, but extracting and using emotional cues from audio poses challenges. This paper introduces MFHCA, a novel method for Speech Emotion Recognition using Multi-Spatial Fusion and Hierarchical Cooperative Attention on spectrograms and raw audio. We employ the Multi-Spatial Fusion module (MF) to efficiently identify emotion-related spectrogram regions and integrate Hubert features for higher-level acoustic information. Our approach also includes a Hierarchical Cooperative Attention module (HCA) to merge features from various auditory levels. We evaluate our method on the IEMOCAP dataset and achieve 2.6\% and 1.87\% improvements on the weighted accuracy and unweighted accuracy, respectively. Extensive experiments demonstrate the effectiveness of the proposed method.
\end{abstract}


\section{Introduction}
Speech emotion recognition (SER) technology plays a crucial role in intelligent human-machine interaction systems~\cite{c1}, as it can identify the speaker's emotional state, thereby enhancing the naturalness of human-machine interaction. Recently, some multimodal SER methods~\cite{c2,c3,c22} have achieved significantly higher accuracy than speech emotion recognition. However, the model can only extract information from the speech signal in specific scenarios, such as voice assistants and phone customer service. Therefore, we focus on extracting emotional information from speech signals.

Many SER models can be seen as a combination of deep feature extractors and classifiers in their structure. In order to attain powerful representational capabilities, numerous researchers have enhanced network models. Xu et al.~\cite{c4} proposed an Attention-based CNN (ACNN) to acquire more effective features, aiming to enhance the performance of SER. Liu et al.~\cite{c5} integrated multi-scale CNN with time-frequency CNN, concurrently modeling local and global information. Chen et al.~\cite{c6} employed a dynamic window transformer to locate significant regions at different time scales, capturing essential information.
Chen et al.~\cite{c7} devised a Deformable Speech Transformer (DST) that captures multi-granular emotional information through deformable attention windows. Although these methods have achieved higher performance, the intricate structures pose challenges for computational resources.

The outstanding performance of speech self-supervised learning in downstream tasks such as automatic speech recognition (ASR) has opened up new avenues for developing SER. Xia et al.~\cite{c10} fine-tuned the Wav2Vec model~\cite{c11}, leveraging learned features for SER tasks. He et al.~\cite{c8} designed a SER method based on the cross-attention transformer that integrates three acoustic features, one of which is extracted from the Wav2Vec2 model~\cite{c9}. Gat I et al.~\cite{c12} proposed a speaker feature normalization framework based on self-supervised feature representation, achieving outstanding performance in SER.

In this paper, we propose a simple SER method (MFHCA) based on Multi-Spatial Fusion module (MF) and Hierarchical Cooperative Attention module (HCA), the MF is primarily composed of several Global Receptive Field block (GRF). Unlike~\cite{c13,c8}, which uses many acoustic features for SER, we only use features from the Hubert model~\cite{c14} and log Mel spectrogram. The Hubert model is trained by performing K-means clustering on MFCC or Hubert features, serving as the training objective. This approach enables the learning of rich feature representations. The log Mel spectrogram better aligns with the auditory characteristics of the human ear. In MFHCA, we employ the MF to extract features in different scale spaces along the temporal and frequency directions, enhancing the model's focus on emotion-related features. Subsequently, the two sets of features are cascaded using the HCA. Finally, a classifier consisting of three fully connected layers is employed to classify the features, completing the emotion recognition task.

Our primary contributions can be summarized as follows:
\begin{itemize}
\item We propose a new SER network that combines joint self-supervised features and the log Mel spectrogram. We alse design a Hierarchical Cooperative Attention module (HCA) to integrate the two sets of features interactively.
\item We propose a novel spectrum-based lightweight feature extraction module, denoted as Multi-Spatial Fusion module (MF), which captures dependencies and positional information in different scale spaces, aiding the network in locating emotional information.
\end{itemize}

\begin{figure*}[t]
\begin{minipage}[b]{1.0\linewidth}
  \centering
  \centerline{\includegraphics[width=18cm]{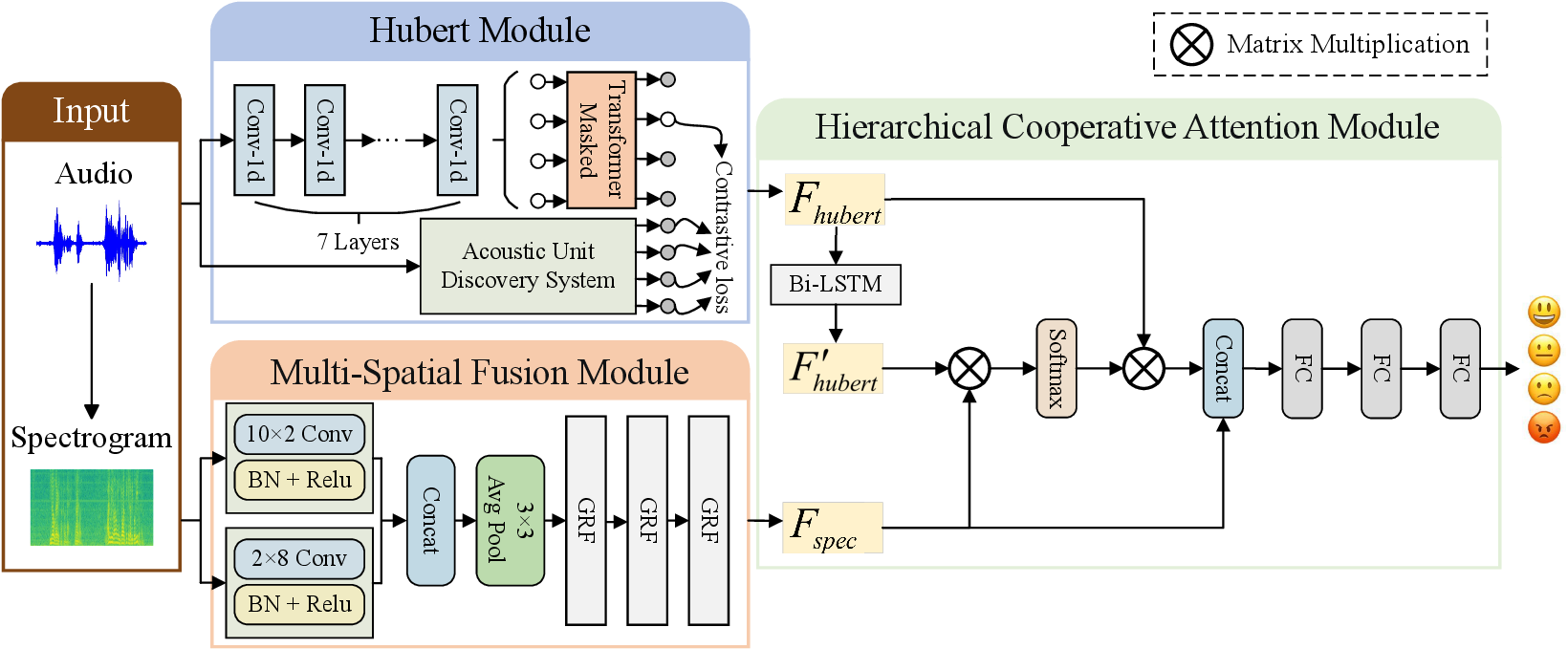}}
  \caption{The overall architecture of our proposed method.}\medskip
  \label{fig:model}
\end{minipage}
\end{figure*}

\section{Proposed method}

This section provides an overview of the proposed SER method. In the following subsections, we provide a detailed description of MF, GRF, and HCA.

\subsection{Overall Architecture}

The overall structure of the end-to-end emotion recognition method we proposed is illustrated in Fig.\ref{fig:model}. The method employs two parallel encoders to extract features from the log Mel spectrogram and raw audio. Specifically, MF uses two parallel convolutional layers to capture low-level features from the log Mel spectrogram in both temporal and frequency directions. GRF extracts dependencies and positional information from the features, enhancing the model's ability to learn emotion-related features. In order to save computational time, we utilize Hubert as a feature extractor to obtain feature sequences from the audio directly without fine-tuning Hubert~\cite{c12}. HCA hierarchically integrates the two sets of features, and a classifier composed of three fully connected layers utilizes the concatenated features for emotion recognition.

\begin{figure*}[htb]
  \centering
  \centerline{\includegraphics[width=15cm]{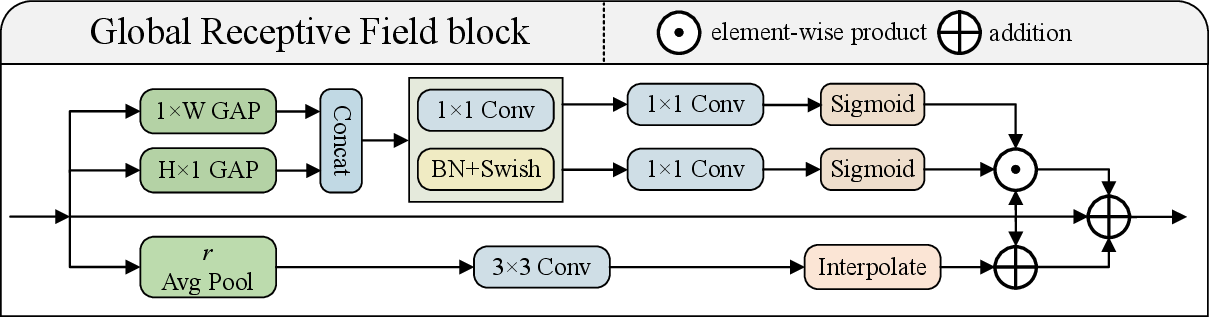}}
  \caption{Global Receptive Field block}
  \label{fig:grf}
\end{figure*}

\subsection{Multi-Spatial Fusion Module}

Fig. \ref{fig:model} illustrates the structure of MF, which is primarily composed of a parallel convolutional layer, a pooling layer, and three Global Receptive Field blocks (GRF). We use processed log Mel spectrogram as input. For the first layer of the module, we employ two parallel convolutions with kernel sizes of (10, 2) and (2, 8), respectively, to extract features in the time and frequency directions. Not all information in the input features is directly relevant to the SER task. Accurately localizing emotional information enables a more comprehensive representation of emotions. To address this problem, we propose a simple block called GRF, and its structure is illustrated in Fig. \ref{fig:grf}.

GRF consists of three parts. Specifically, the first part represents the input features, denoted by $X$. Global pooling can capture spatial information, but compressing and preserving global spatial information through squeezing may lead to losing positional information. In the second part, we adopt encoding in both the temporal and frequency directions as a replacement for average pooling. We employ global average pooling(GAP) with kernels $(H, 1)$ and $(1, W)$ to obtain a set of encodings with dependency and positional information. The output for the $c$-th channel at height $h$ is represented as $z_{c}(h)$, and the output for the $c$-th channel at width $w$ is denoted as $z_{c}(w)$. 

\begin{equation}
z_{c}(h)=\frac{1}{W}\sum_{0 \textless i \leq W} x_{c}(h,i),\label{eq1}
\end{equation}

\begin{equation}
z_{c}(w)=\frac{1}{H}\sum_{0 \textless j \leq H} x_{c}(j,w).\label{eq2}
\end{equation}

We utilize 1×1 convolutions to interactively pass positional information between $z_{c}(h)$ and $z_{c}(w)$ to guide the model in locating regions relevant to emotional information. Additionally, this allows for learning relationships between channels.

\begin{equation}
f=\delta(BN(F([z(h),z(w)^{T}]))),\label{eq3}
\end{equation}

where, $[z(h),z(w)^{T}]$ represents a cascading operation in the spatial dimension, $F$ denotes a 1×1 convolution operation, $BN$ stands for Batch Normalization, $\delta$ means the non-linear activation function $swish$. Subsequently, we partition feature $f$ in the spatial dimension into two independent tensors, $f_{h}$ and $f_{w}$.

\begin{equation}
g_{h}=\sigma(F_{h}(f_{h})),\label{eq4}
\end{equation}

\begin{equation}
g_{w}=\sigma(F_{w}(f_{w})),\label{eq5}
\end{equation}

here, $F_{h}$ and $F_{w}$ represent 1×1 convolutions, and their role is to adjust the number of channels in $f_{h}$ and $f_{w}$ to be consistent with the input $X$. $\sigma$ denotes the sigmoid function. Finally, the outputs of these two parts are used as attention weights applied to the input $X$. $Y_{a}$ represents the final output.
\begin{equation}
Y_{a}(i,j)=x(i,j)  \odot  g_{h}(i)  \odot  g_{w}(j),\label{eq6}
\end{equation}
where, $\odot$ represents hadamard product.

In the second part, We employ average pooling to perform downsampling on the input with a rate of $r$. Assuming the input is a 3D tensor with a shape of $(C, H, W)$, the shape becomes $(C, H/r, W/r)$ after pooling. Equation (\ref{eq7}) demonstrates the computation process.
\begin{equation}
Y_{b}=X \oplus Up(F(AvgPool_{r}(X))),\label{eq7}
\end{equation}
where $AvgPool_{r}$ represents average pooling, $F$ is a convolutional transformation, and $Up$ is a bilinear interpolation operator used to restore features from a smaller-scale space to the original feature space. $\oplus$ is addition. $Y_{b}$ is the output. The features are transformed into low-dimensional embeddings through downsampling. After transformation by convolutional layers, the low-dimensional embeddings encompass a larger receptive field, enabling the calibration of the convolutional layers in $Y_{a}$. This communication between the convolutional layers in $Y_{a}$ and $Y_{b}$ expands the receptive field in spatial positions. Each spatial position can also gather context from its surroundings, and dependencies between channels can be learned. Benefiting from this internal information exchange, GRF can generate more discriminative representations. The output of GRF is denoted as Y.
\begin{equation}
Y=X \oplus Y_{a} \oplus Y_{b},\label{eq8}
\end{equation}
where $\oplus$ represents addition.

\subsection{Hubert Module}
Hubert is a transformer-based model composed of three components: a waveform encoder, an Acoustic Unit Discovery System, and a Bert encoder. The Acoustic Unit Discovery System is an unsupervised clustering model trained to label input speech and generate discrete target sequences. The waveform encoder consists of seven layers of convolutional networks, extracting low-level features. The Bert encoder comprises multiple identical Transformer blocks, enabling it to learn contextualized representations.

Hubert conducts k-means and GMM clustering on MFCC features to discretize speech signals, creating training targets. This discretized target sequence derived from MFCC allows the model to learn more representations, such as vocabulary information related to speech recognition and features associated with speakers and emotions. This diversity of information gives it a distinct advantage in emotion recognition. The pre-trained HuBERT model excels at capturing richer, multi-layered speech information, leading to enhanced generalization performance in speech emotion recognition.

\subsection{Hierarchical Cooperative Attention Module}

Research has shown that the hierarchical fusion of different features is helpful in SER tasks~\cite{c19}. We employ the Hubert model as a feature extractor without fine-tuning, and the learned features contain rich information. In addition to emotion recognition-related information, information is also relevant to other downstream tasks. Considering these aspects, we designed the HCA module. Spectrogram features are frame-level, and Hubert features contain contextual information. MF can pinpoint the location of emotional information. We use features from MF as guidance for Hubert features, assisting the network in focusing on emotional information within Hubert features and jointly hierarchically obtaining more powerful feature representations. We use $f_{hubert}$ to denote the output of the hidden layer from Hubert, $f_{hubert}'$ to represent the output of Hubert features after passing through a BiLSTM, and $f_{spec}$ to denote the spectrogram. We perform the following computations:
\begin{equation}
f_{att}= softmax(f_{spec} \otimes f_{hubert}'),\label{eq9}
\end{equation}
\begin{equation}
f_{att}'= f_{hubert} \otimes f_{att},\label{eq10}
\end{equation}
\begin{equation}
f_{att}''= concat(f_{spec},f_{att}'),\label{eq11}
\end{equation}
where $\otimes$ represents Matrix multiplication, we append a BiLSTM to the end of the Hubert model to obtain a temporal sequence $f_{hubert}'$ containing contextual information. A co-attention operation is performed between $f_{hubert}'$ and $f_{spec}$ to obtain weights $f_{att}$. Then, a second-level operation is conducted between $f_{att}$ and $f_{hubert}$, resulting in $f_{att}'$. At this stage, the emotional features in $f_{att}'$ are enhanced. Finally, $f_{att}'$ is concatenated with $f_{spec}$ to obtain the final feature representation. The emotion classification is achieved through a classifier consisting of three fully connected layers.

\section{EXPERIMENTS AND DISCUSSION}

\subsection{Dataset and Implementation Details}

We evaluated our approach on the Interactive Emotional dyadic Capture database (IEMOCAP) \cite{c15}, which consists of 10 actors engaged in 5 dyadic sessions, each featuring a unique pair of male and female actors. Following the work of others \cite{c16,c17,c18}, we merged $happy$ and $excited$ into a single category labeled as $happy$, while also considering $neutral$, $sad$, $happy$, and $angry$ emotions.

In the preprocessing stage, to facilitate comparison with baseline methods \cite{c13}, we also segment the original audio signal into 3-second-long segments. We apply zero padding when the audio segment is less than 3 seconds. Spectrograms are extracted using a Hamming window with a window length of 40ms and a window shift of 10ms. Each windowed block is treated as a frame, and a Discrete Fourier Transform (DFT) of length 800 is applied to transform each frame into the frequency domain. The first 200 DFT points are used as input spectrogram features. Hubert features correspond to the output of the final hidden layer of the Hubert model. 

The model optimizer is Adam, the learning rate is set with $1 \times 10^{-5}$, the training batch size is 32, and early stopping is configured for 10 epochs, the optimization function is Cross Entropy Loss. We employed a 10-fold leave-one-speaker-out cross-validation strategy to assess the model's performance, with evaluation metrics being Unweighted Accuracy (\textbf{UA}) and Weighted Accuracy (\textbf{WA}).

\subsection{Results and Comparison}

\begin{table}[b]
\centering
\tabcolsep=0.3cm
\caption{Comparison with known state-of-the-art systems on IEMOCAP, where A and T denote audio and text modalities.}
\renewcommand\arraystretch{1.2}
\label{tab1}
\scalebox{1}{
\begin{tabular}{l|l|ll}
\hline
\textbf{Model}  & \textbf{Modality}  & \textbf{WA($\uparrow$)} & \textbf{UA($\uparrow$)} \\ \hline
HNSD \cite{c18}     & A            & 70.50         & 72.50         \\
Co-attention \cite{c13}  & A    & 71.64         & 72.70         \\
SMW\_CAT \cite{c8}  & A    & 73.80         & 74.25         \\
E2e ASR \cite{c20}    & A+T          & 71.70         & 72.60         \\
Cross-representation \cite{c21} & A+T & 73.00         & 73.50         \\
\textbf{MFHCA(ours)}           & A               & \textbf{74.24}         & \textbf{74.57}         \\ \hline
\end{tabular}
}
\end{table}

Table 1 compares our proposed method and the baselines on WA and UA. We compare our method with these approaches under two speaker-independent validation strategies. The experimental results demonstrate that our approach achieves the best performance. HNSD~\cite{c18} extracts static and dynamic features for emotion recognition from the log-Mel filter bank feature. The model can achieve better performance when utilizing features from the pre-trained model~\cite{c20}. In order to obtain suitable emotion representations, SER systems based on pre-trained models incorporate multiple speech features using methods such as co-attention-based approaches~\cite{c13}, cross-representation learning~\cite{c21}, and cross-attention transformers~\cite{c8}. Moreover, our model has 54.26\% fewer parameters than the baseline~\cite{c13} while achieving higher performance. The experimental results demonstrate the effectiveness and lightweight of our method.

\subsection{Ablation Study}

\begin{table}[t]
\centering
\tabcolsep=0.4cm
\caption{Ablation study on the proposed model.}
\renewcommand\arraystretch{1.2}
\label{tab2}
\scalebox{1}{
\begin{tabular}{l|l|l|l|l}
\hline
\textbf{Models}    & \textbf{MF} & \textbf{HCA} & \textbf{WA($\uparrow$)}    & \textbf{UA($\uparrow$)}    \\ \hline
\multirow{2}{*}{Spec}        &  \makecell[c]{\ding{55}}   & \makecell[c]{\ding{55}}     & 62.13 & 62.25 \\
                             &  \makecell[c]{\usym{1F5F8}}   & \makecell[c]{\ding{55}}     & 62.21 & 62.37 \\ \hline
Hubert                 & \makecell[c]{\ding{55}}    & \makecell[c]{\ding{55}}      & 69.99 & 70.57 \\ \hline
\multirow{4}{*}{Spec+W2V2} & \makecell[c]{\ding{55}}    & \makecell[c]{\ding{55}}      & 70.05 & 71.30 \\
                             & \makecell[c]{\usym{1F5F8}}    & \makecell[c]{\ding{55}}      & 71.73 & 72.32 \\
                             & \makecell[c]{\ding{55}}    & \makecell[c]{\usym{1F5F8}}      & 70.72 & 72.09 \\
                             & \makecell[c]{\usym{1F5F8}}    & \makecell[c]{\usym{1F5F8}}      & \textbf{72.00} & \textbf{73.44} \\ \hline
\multirow{4}{*}{Spec+Hubert} & \makecell[c]{\ding{55}}    & \makecell[c]{\ding{55}}      & 72.13 & 72.51 \\
                             & \makecell[c]{\usym{1F5F8}}    & \makecell[c]{\ding{55}}      & 73.72 & 74.53 \\
                             & \makecell[c]{\ding{55}}    & \makecell[c]{\usym{1F5F8}}      & 73.19 & 73.72 \\
                             & \makecell[c]{\usym{1F5F8}}    & \makecell[c]{\usym{1F5F8}}      & \textbf{74.24} & \textbf{74.57} \\ \hline
\end{tabular}
}
\end{table}

\begin{table}[t]
\centering
\tabcolsep=0.5cm
\caption{The impact of quantity and channel variation of GRF on performance.}
\renewcommand\arraystretch{1.2}
\label{tab3}
\scalebox{1}{
\begin{tabular}{ll|ll}
\hline
\textbf{Num} & \textbf{Channel} & \textbf{WA($\uparrow$)} & \textbf{UA($\uparrow$)} \\ \hline
2            & 16-32            & 73.76       & 74.02       \\
\textbf{3}            & \textbf{16-32-48}         & \textbf{74.24}       & \textbf{74.57}       \\
3            & 16-32-64         & 73.60       & 74.49       \\
4            & 16-32-48-64      & 73.67       & 74.03       \\
4            & 16-32-64-128     & 72.75       & 73.78       \\ \hline
\end{tabular}
}
\end{table}

\begin{table}[t]
\centering
\tabcolsep=0.9cm
\caption{The impact of sampling ratio $r$ on performance.}
\renewcommand\arraystretch{1.2}
\label{tab4}
\scalebox{1}{
\begin{tabular}{lll}
\hline
\multicolumn{1}{l|}{\textbf{$r$}} & \textbf{WA($\uparrow$)} & \textbf{UA($\uparrow$)} \\ \hline
\multicolumn{1}{l|}{1/2}        & 73.82         & 74.33         \\
\multicolumn{1}{l|}{\textbf{1/4}}        & \textbf{74.24}         & \textbf{74.57}         \\
\multicolumn{1}{l|}{1/8}        & 72.22         & 73.34         \\
\multicolumn{1}{l|}{1/16}       & 71.12         & 72.69         \\ \hline
\end{tabular}
}
\end{table}

In Table \ref{tab2}, we conducted ablation experiments to demonstrate the effectiveness of MF and HCA. "Hubert" denotes features extracted using Hubert, and "W2V2" represents features extracted using wav2vec2. Specifically, in the experiment involving two features, the absence of HCA implementation does not imply the exclusion of fusion methods. To ensure comparability, we employed baseline fusion methods. We initially conducted experiments using single features and found that MF contributes to the extraction of emotional information in the spectrogram. The pre-trained model used in the baseline is wav2vec2. For a more intuitive comparison, we conducted experiments based on Spec and W2V2. The results indicate that our proposed method performs better in learning emotional information from features compared to the baseline. Subsequently, we further validated our method on Spec and Hubert, achieving improved recognition accuracy. In Table \ref{tab3}, we employed grid search to investigate the impact of changes in the number of GRFs and channels on the experimental results. In Table \ref{tab4}, we similarly studied the influence of variations in the scaling factor $r$ on the experimental results.

\begin{figure}[t]
\begin{minipage}[b]{.48\linewidth}
  \centering
  \centerline{\includegraphics[width=3.8cm]{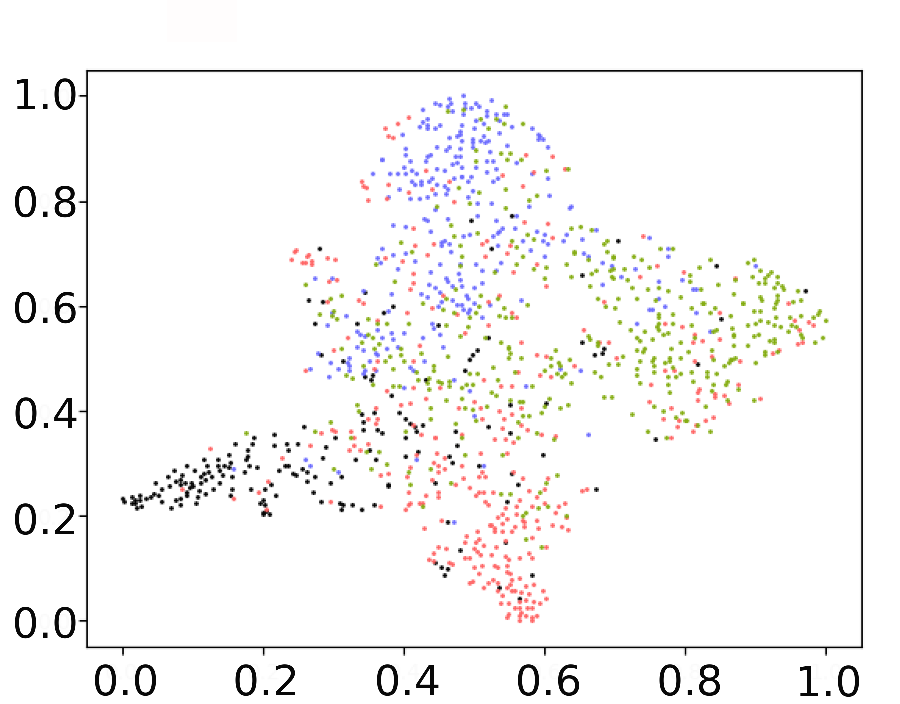}}
  \centerline{(a) Final features w/o HCA}\medskip
\end{minipage}
\hfill
\begin{minipage}[b]{0.48\linewidth}
  \centering
  \centerline{\includegraphics[width=3.8cm]{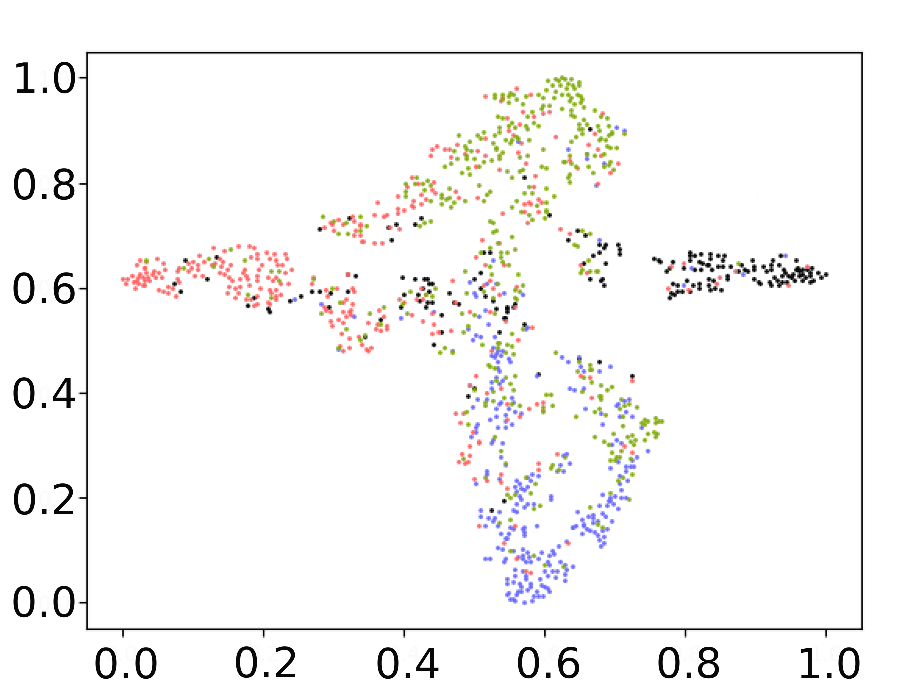}}
  \centerline{(b) Final features w/ HCA}\medskip
\end{minipage}
\caption{The t-SNE visualization of feature distribution. (a) and (b) are the final combined features without and with the proposed HCA.}
\label{fig:sne}
\end{figure}

To further validate our approach, we employed t-SNE visualization to depict the feature distributions with and without HCA in Fig. \ref{fig:sne}. When HCA was used, the dense areas of feature distribution were reduced, and classification boundaries became more distinct. 

\section{CONCLUSION}

This paper presents MFHCA, a novel SER approach utilizing MF for spectrogram feature extraction. Our method capitalizes on GRF's advantages to overcome CNN's limitations in capturing global information. HCA fosters interaction between spectral diagrams and Hubert features, jointly harvesting emotional representations from both components. Experimental results on IEMOCAP substantiate the effectiveness of MF and HCA in our approach.

\section*{Acknowledgment}

The following projects jointly supported this work: the Tianshan Excellence Program Project of Xinjiang Uygur Autonomous Region, China (2022TSYCLJ0036); the Central Government Guides Local Science and Technology Development Fund Projects (ZYYD2022C19); the National Natural Science Foundation of China under Grant 62303259.


\begin{thebibliography}{00}
\bibitem{c1} Cowie R, Douglas-Cowie E, Tsapatsoulis N, et al. Emotion recognition in human-computer interaction[J]. IEEE Signal processing magazine, 2001, 18(1): 32-80.
\bibitem{c2} Chen W, Xing X, Xu X, et al. Key-sparse transformer for multimodal speech emotion recognition[C]//ICASSP 2022-2022 IEEE International Conference on Acoustics, Speech and Signal Processing (ICASSP). IEEE, 2022: 6897-6901.
\bibitem{c3} Wang S, Ma Y, Ding Y. Exploring complementary features in multi-modal speech emotion recognition[C]//ICASSP 2023-2023 IEEE International Conference on Acoustics, Speech and Signal Processing (ICASSP). IEEE, 2023: 1-5.
\bibitem{c4} Xu M, Zhang F, Khan S U. Improve accuracy of speech emotion recognition with attention head fusion[C]//2020 10th annual computing and communication workshop and conference (CCWC). IEEE, 2020: 1058-1064.
\bibitem{c5} Liu J, Liu Z, Wang L, et al. Speech emotion recognition with local-global aware deep representation learning[C]//ICASSP 2020-2020 IEEE International Conference on Acoustics, Speech and Signal Processing (ICASSP). IEEE, 2020: 7174-7178.
\bibitem{c6} Chen S, Xing X, Zhang W, et al. DWFormer: Dynamic Window Transformer for Speech Emotion Recognition[C]//ICASSP 2023-2023 IEEE International Conference on Acoustics, Speech and Signal Processing (ICASSP). IEEE, 2023: 1-5.
\bibitem{c7} Chen W, Xing X, Xu X, et al. DST: Deformable speech transformer for emotion recognition[C]//ICASSP 2023-2023 IEEE International Conference on Acoustics, Speech and Signal Processing (ICASSP). IEEE, 2023: 1-5.
\bibitem{c8} He Y, Minematsu N, Saito D. Multiple acoustic features speech emotion recognition using cross-attention transformer[C]//ICASSP 2023-2023 IEEE International Conference on Acoustics, Speech and Signal Processing (ICASSP). IEEE, 2023: 1-5.
\bibitem{c9} Baevski A, Zhou Y, Mohamed A, et al. wav2vec 2.0: A framework for self-supervised learning of speech representations[J]. Advances in neural information processing systems, 2020, 33: 12449-12460.
\bibitem{c10} Xia Y, Chen L W, Rudnicky A, et al. Temporal Context in Speech Emotion Recognition[C]//Interspeech. 2021, 2021: 3370-3374.
\bibitem{c11} Schneider S, Baevski A, Collobert R, et al. wav2vec: Unsupervised pre-training for speech recognition[J]. arXiv preprint arXiv:1904.05862, 2019.
\bibitem{c12} Gat I, Aronowitz H, Zhu W, et al. Speaker normalization for self-supervised speech emotion recognition[C]//ICASSP 2022-2022 IEEE International Conference on Acoustics, Speech and Signal Processing (ICASSP). IEEE, 2022: 7342-7346.
\bibitem{c13} Zou H, Si Y, Chen C, et al. Speech emotion recognition with co-attention based multi-level acoustic information[C]//ICASSP 2022-2022 IEEE International Conference on Acoustics, Speech and Signal Processing (ICASSP). IEEE, 2022: 7367-7371.
\bibitem{c14} Hsu W N, Bolte B, Tsai Y H H, et al. Hubert: Self-supervised speech representation learning by masked prediction of hidden units[J]. IEEE/ACM Transactions on Audio, Speech, and Language Processing, 2021, 29: 3451-3460.
\bibitem{c15} Busso C, Bulut M, Lee C C, et al. IEMOCAP: Interactive emotional dyadic motion capture database[J]. Language resources and evaluation, 2008, 42: 335-359.
\bibitem{c16} Guo L, Wang L, Xu C, et al. Representation learning with spectro-temporal-channel attention for speech emotion recognition[C]//ICASSP 2021-2021 IEEE International Conference on Acoustics, Speech and Signal Processing (ICASSP). IEEE, 2021: 6304-6308.
\bibitem{c17} Wu W, Zhang C, Woodland P C. Emotion recognition by fusing time synchronous and time asynchronous representations[C]//ICASSP 2021-2021 IEEE International Conference on Acoustics, Speech and Signal Processing (ICASSP). IEEE, 2021: 6269-6273.
\bibitem{c18} Cao Q, Hou M, Chen B, et al. Hierarchical network based on the fusion of static and dynamic features for speech emotion recognition[C]//ICASSP 2021-2021 IEEE International Conference on Acoustics, Speech and Signal Processing (ICASSP). IEEE, 2021: 6334-6338.
\bibitem{c19} Tian L, Moore J, Lai C. Recognizing emotions in spoken dialogue with hierarchically fused acoustic and lexical features[C]//2016 IEEE Spoken Language Technology Workshop (SLT). IEEE, 2016: 565-572.
\bibitem{c20} Lu Z, Cao L, Zhang Y, et al. Speech sentiment analysis via pre-trained features from end-to-end asr models[C]//ICASSP 2020-2020 IEEE International Conference on Acoustics, Speech and Signal Processing (ICASSP). IEEE, 2020: 7149-7153.
\bibitem{c21} Makiuchi M R, Uto K, Shinoda K. Multimodal emotion recognition with high-level speech and text features[C]//2021 IEEE Automatic Speech Recognition and Understanding Workshop (ASRU). IEEE, 2021: 350-357.
\bibitem{c22} Yu Y, Jia Z, Shi F, et al. WeaveNet: End-to-End Audiovisual Sentiment Analysis[C]//International Conference on Cognitive Systems and Signal Processing. Singapore: Springer Nature Singapore, 2021: 3-16.
\end{thebibliography}
\end{document}